\documentclass[sigplan,screen]{acmart}

\copyrightyear{2027}
\acmYear{2027}
\setcopyright{cc}
\setcctype{by}
\acmConference[ASPLOS '27]{Proceedings of the 32nd ACM International Conference on Architectural Support for Programming Languages and Operating Systems, Volume 1}{April 11--15, 2027}{Heraklion, Greece.}
\acmBooktitle{Proceedings of the 32nd ACM International Conference on Architectural Support for Programming Languages and Operating Systems, Volume 1 (ASPLOS '27), April 11--15, 2027, Heraklion, Greece}
\acmISBN{979-8-4007-3012-2/2027/04}
\acmDOI{10.1145/3845814.3850033}
\usepackage{algorithm}
\usepackage{algpseudocode}
\usepackage{subcaption}
\usepackage{pifont}
\usepackage{paralist}
\usepackage{enumitem}
\usepackage{balance}

\begin{document}

\title{FORGE: Procedural Weight Generation for Sparse MoE Inference}

\author{Chanyoung Moon}
\affiliation{
  \institution{University of Wisconsin-Madison}
  \city{Madison}
  \state{Wisconsin}
  \country{USA}}
\affiliation{
  \institution{Korea University}
  \city{Seoul}
  \country{Republic of Korea}}
\email{rchan02@korea.ac.kr}

\author{Travis Pence}
\affiliation{
  \institution{University of Wisconsin-Madison}
  \city{Madison}
  \state{Wisconsin}
  \country{USA}}
\email{tnpence@wisc.edu}

\author{Daisuke Yamada}
\affiliation{
  \institution{University of Wisconsin-Madison}
  \city{Madison}
  \state{Wisconsin}
  \country{USA}}
\email{dyamada2@wisc.edu}

\author{Jiaqi Mo}
\affiliation{
  \institution{University of Wisconsin-Madison}
  \city{Madison}
  \state{Wisconsin}
  \country{USA}}
\email{mo29@wisc.edu}

\author{Yanxin Zhang}
\affiliation{
  \institution{University of Wisconsin-Madison}
  \city{Madison}
  \state{Wisconsin}
  \country{USA}}
\affiliation{
  \institution{NVIDIA}
  \city{Santa Clara}
  \state{California}
  \country{USA}}
\email{yanxinz@nvidia.com}

\author{Vikas Singh}
\affiliation{
  \institution{University of Wisconsin-Madison}
  \city{Madison}
  \state{Wisconsin}
  \country{USA}}
\email{vsingh@biostat.wisc.edu}

\author{Karthikeyan Sankaralingam}
\affiliation{
  \institution{University of Wisconsin-Madison}
  \city{Madison}
  \state{Wisconsin}
  \country{USA}}
\affiliation{
  \institution{NVIDIA}
  \city{Santa Clara}
  \state{California}
  \country{USA}}
\email{karu@cs.wisc.edu}

\renewcommand{\shortauthors}{Moon et al.}

\begin{abstract}
We introduce \emph{FORGE} (Fast On-chip Rotor Generation for Experts), a {\em procedural} weight generation system for Sparse Mixture-of-Experts (MoE) inference. Rather than storing and loading every expert weight tensor from HBM, we show how a compact accelerator inside each Streaming Multiprocessor can regenerate them {\em  on-the-fly} from $O(\log^2 d)$ rotor parameters per projection. Per replaced layer, gigabytes of stored expert weights are approximated with under a megabyte of seeds.
This addresses the dual memory bottleneck of autoregressive Sparse MoE decode, where per-expert FFN operations are HBM-bandwidth-bound at small routed-token counts and the aggregate expert footprint forces multi-node Expert Parallelism (EP) with all-to-all traffic. We realize the approach as the Rotor Generation Unit (RGU), a fixed-function datapath that generates FFN weight tiles without competing with Tensor Core execution.
With $30\%$ of the MoE FFN layers replaced across five Sparse MoE models, we estimate $1.3$--$2.8\times$ higher system throughput at matched GPU count and $1.10$--$1.15\times$ lower per-user decode latency at short context. In exchange, downstream accuracy on three model families falls below that of the same model fine-tuned without rotor replacement, and the replacement ratio controls this tradeoff. FORGE also expands KV cache capacity by $1.2$--$2.6\times$. It is agnostic to the bit width at which weights are stored, so it can compose with quantization. The RGU is fully implemented/verified in SystemVerilog at ${\sim}$0.9\% SM area overhead (0.032~mm\textsuperscript{2}/SM, 7nm).
\end{abstract}




\maketitle

\section{Introduction}

Short-context autoregressive decode, e.g., agentic tool-use chains, short-prompt chat, and single-user serving, is the regime where per-token latency is bottlenecked by the cost of moving weights out of HBM. This is because each decode step produces a single token, leading FFN projections to collapse to low-arithmetic-intensity operations (GEMVs) whose wall-clock is set by weight transfer rather than arithmetic~\cite{pope2023efficient}. For long context or large batches, other bottlenecks (KV cache, Attention FLOPs) take over.

There is broad consensus that sparse Mixture-of-Experts (MoE) architectures~\cite{shazeer2017moe, switchtransformer, gshard} are now the dominant design for frontier open-weight LLMs. It turns out that this architecture amplifies the bottlenecks of short-context decode along three distinct axes. All three lead back to one architectural fact: expert weights are the overwhelming majority of the parameter footprint, i.e., 94--98\% across the five frontier MoEs in Table~\ref{tab:model_configs}, and constantly occupy HBM capacity.

\begin{figure}[!t]
    \centering
    \includegraphics[width=\linewidth]{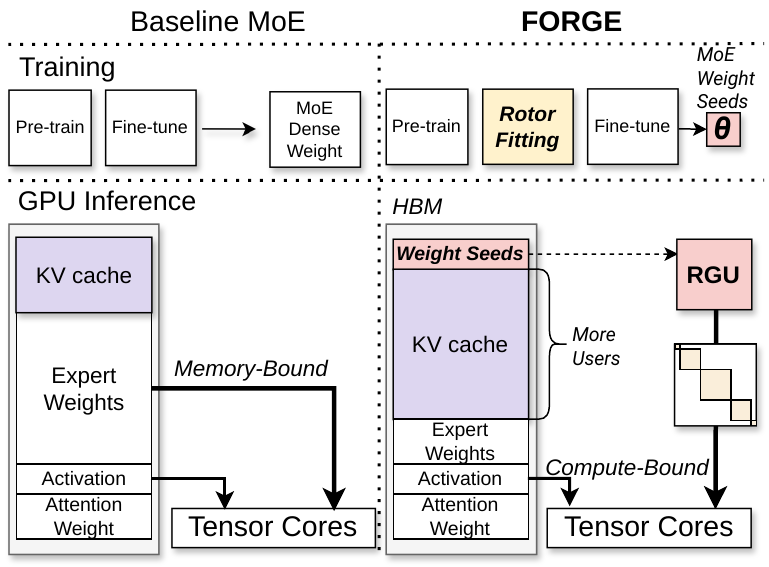}
    \caption{FORGE replaces HBM-resident expert weights with on-chip procedural regeneration. \textbf{(Top)} Training adds an offline rotor-fitting step that compresses each dense expert into a rotor-parameter seed $\theta$. \textbf{(Bottom)} At inference, the RGU inside each SM materializes weight tiles on-the-fly from seeds alone, reclaiming HBM capacity for an expanded KV cache.}
    \label{fig:teaser}
\end{figure}

\textbf{(P1) HBM bandwidth bottleneck at small effective batch.} We know that MoE top-$K$ routing assigns each token to $K$ of $E$ total experts, e.g., $K{=}8$ of $E{=}256$ for DeepSeek-v3.2~\cite{deepseekv32}; $K{=}10$ of $E{=}512$ for Qwen3-Coder-Next~\cite{qwen3coder}. But a serving batch of $B$ tokens therefore yields only $BK/E$ tokens per expert under uniform routing. At $B{=}32$ on DeepSeek-v3.2, each expert sees a single token, collapsing its Up/Gate/Down projections to GEMVs. Per-token decode latency then becomes dominated by HBM weight transfer, leaving the Tensor Cores idle~\cite{pope2023efficient}.

\textbf{(P2) HBM capacity bottleneck.} At FP8, total MoE weight footprints range from 31GB (Qwen3-30B~\cite{qwen3}) to 742GB (GLM-5~\cite{glm5}). All but the smallest exceed the 80GB HBM of a single H100. While offloading weights to host memory is an alternative, it severely bottlenecks decode performance to PCIe bandwidth limits~\cite{liminal}. This forces Expert Parallelism (EP)~\cite{gshard, deepspeedmoe, tutel} across multiple GPUs and across multiple nodes for large models---paying a cross-node all-to-all for every decode step. Now, within each GPU, weights also consume HBM that would otherwise serve KV cache, capping the number of concurrent users per GPU. The capacity problem thus has two consequences: deployment topology (how many GPUs, how much interconnect) and serving concurrency (how many users per GPU).

\textbf{(P3) Compression does not change the access pattern.} Quantization~\cite{fp8formats, awq, gptq} (FP8 on Hopper, INT4 variants) is the standard response to address both pressures. It does shrink per-parameter bytes, but the GPU must still fetch every activated weight from HBM on each decode step and reserve HBM capacity for the weights. Quantization thus reduces the bandwidth bottleneck and lowers the capacity ceiling, but we observe it eliminates neither. More aggressive schemes (e.g., INT4) push this further along the same axis without changing the underlying access pattern.

We observe that (P1), (P2), and (P3) all derive from a single architectural assumption that is rarely challenged: expert weights are a {\em static payload} resident in HBM. 

\textbf{Our approach.} The main idea is to instead treat each expert projection as {\em approximately procedurally synthesizable} (i.e., the end product of a recipe rather than a stored tensor) so that, in principle, all three problems can be addressed {\em at once}.  Concretely, recent work by Pence et al.~\cite{pence2025composing} on a Clifford-algebra object, rotors, offers the theoretical feasibility: $O(d^2)$ linear projections can be approximated by $O(\log^2 d)$ rotor parameters. Operationally, those parameters are angles, and the projection is rebuilt by applying the corresponding rotations in a fixed sequence.

We take this, with appropriate adjustments, from a theoretical formalism to hardware. Under our construction, each pre-trained expert in a replaced MoE layer is fit offline to a rotor-parameter seed, and at inference the seed stands in for the dense expert tensor for that layer. A fixed-function \textbf{Rotor Generation Unit (RGU)} inside each Streaming Multiprocessor materializes weight tiles {\em on demand} for those to-be-replaced layers and writes them directly into Shared Memory, feeding the Tensor Core pipeline without an HBM weight load on the replaced path (Figure~\ref{fig:teaser}). 

Against (P1), on replaced layers this swaps the HBM weight-loading path for a dedicated on-chip generator, shifting the per-expert MoE FFN kernel from bandwidth-bound to compute-bound. Against (P2), replaced-layer expert storage collapses by more than three orders of magnitude, to under a megabyte of parameter seeds per layer (Table~\ref{tab:model_configs}), freeing close to a third of the weight HBM for expanded KV cache and shrinking the minimum-fit GPU count for the largest models enough to eliminate cross-node deployment. Against (P3), the gains are orthogonal to quantization, as we remove the transfer rather than compress it. The net effect is higher system throughput without giving up per-user latency, which normally trade off against each other, at a cost in downstream accuracy. The substitution is tunable. With $30\%$ of the MoE FFN layers replaced, accuracy falls below that of the same model fine-tuned without rotor replacement (Section~\ref{sec:quality_validation}); higher ratios promise larger system gains (Figure~\ref{fig:replace_ablation}), and keeping accuracy at those ratios is future work.

\emph{Scope.} The throughput gain holds across context lengths, since it rests on freed HBM capacity (Section~\ref{sec:system_impact}). The per-user latency gain is largest at short context and narrows to $1.07$--$1.11\times$ at $T{=}128$K, where KV cache and attention dominate execution time (Section~\ref{sec:decode_latency}). Specifically, our contributions are as follows.

\textbf{Contributions.} \textbf{(a)} We adapt the Clifford-algebra rotor framework~\cite{pence2025composing} from attention layers to MoE FFN decode, and develop a parallel execution model (column, row-pair, and $k$-block independence with precomputed conflict-free pair scheduling) suited to hardware realization. \textbf{(b)} We identify why a software-only path fails: SIMT-based generation saturates instruction issue and incurs pervasive shared-memory bank conflicts, reducing Tensor Core utilization. \textbf{(c)} We design, implement in SystemVerilog, and verify the Rotor Generation Unit, a $0.032\,mm^2$/SM ($\sim$0.9\% of a modern SM~\cite{locuza_ada}; 7\,nm ASAP7 + CACTI) fixed-function accelerator that generates weight tiles directly into Shared Memory. \textbf{(d)} We evaluate five Sparse MoE models with $30\%$ of the MoE FFN layers replaced and measure accuracy on three model families (Section~\ref{sec:quality_validation}). At this point we estimate $1.3$--$2.8\times$ higher system throughput at matched GPU count and $1.10$--$1.15\times$ lower per-user decode latency at short context, at a cost in downstream accuracy relative to the same model fine-tuned without rotor replacement. \textbf{(e)} The gains rest on measured and RTL-backed inputs: a $2.3$--$3.2\times$ per-expert FP8 kernel speedup independent of $r$, $1.2$--$2.6\times$ more KV cache capacity, and a $28$--$29\%$ reduction in total parameter memory.

The remainder of this paper is organized as follows. Section~\ref{sec:background} provides background on Sparse MoE bottlenecks and introduces the rotor projection framework. Section~\ref{sec:rotor_generation} develops the parallel execution model and the rotor generation architecture. Section~\ref{sec:microarch} details the hardware microarchitecture of the RGU. Section~\ref{sec:evaluation} evaluates the proposed design. Section~\ref{sec:related_work} discusses related work, and Section~\ref{sec:conclusion} concludes.

\section{Background and Motivation}
\label{sec:background}

\subsection{Sparse MoE Architecture}

Sparse Mixture-of-Experts (MoE) architectures~\cite{shazeer2017moe, switchtransformer, gshard} replace the single dense FFN in each transformer block with a set of $E$ expert FFN modules and a learned gating network. At each MoE layer, the gating network routes each token to a small subset of experts---typically 8 out of 256 (DeepSeek-v3.2~\cite{deepseekv32}, GLM-5~\cite{glm5}) or 10 out of 512 (Qwen3-Coder-Next~\cite{qwen3coder}). Each activated expert executes a standard SwiGLU FFN~\cite{swiglu}: three linear projections (Up, Gate, Down) mapping between the model's hidden dimension $d$ and the per-expert intermediate dimension $d_{moe}$. Only the activated experts contribute to the forward pass, keeping per-token FLOPs comparable to a dense model of equivalent depth.

While each expert is compact (small $d_{moe}$), the total expert count drives aggregate parameter volume: MoE weight parameters constitute the dominant share of total model footprint across frontier architectures. Table~\ref{tab:model_configs} summarizes the five target models evaluated in this work. At FP8 precision, MoE parameters account for 94--98\% of total model size, ranging from 29~GB (Qwen3-30B~\cite{qwen3}, 128 experts) to 728~GB (GLM-5, 256 experts across 75 MoE layers). This parameter concentration is the source of both the bandwidth and capacity bottlenecks analyzed in Section~\ref{sec:inference_bottleneck}.

\begin{table}[h]
\centering
\caption{Target Sparse MoE model configurations. $E$, $K$: total and activated experts; Seed: the FORGE parameters that replace one layer's experts.}
\label{tab:model_configs}
\small
\begin{tabular}{@{}lrrrrr@{}}
\toprule
Model & $E$ & $K$ & MoE\% & Seed/layer & Comp. \\
\midrule
DeepSeek-v3.2 & 256 & 8 & 94.5 & 594\,KB & $19{,}065\times$ \\
GLM-5 & 256 & 8 & 98.2 & 509\,KB & $19{,}065\times$ \\
Qwen3-235B & 128 & 8 & 96.6 & 276\,KB & $8{,}738\times$ \\
Qwen3-30B & 128 & 8 & 95.0 & 221\,KB & $2{,}731\times$ \\
Qwen3-Coder-Next & 512 & 10 & 98.0 & 776\,KB & $2{,}081\times$ \\
\bottomrule
\end{tabular}
\end{table}

\subsection{Inference Bottleneck and Limits of Compression}
\label{sec:inference_bottleneck}

In autoregressive decode, each token generates a single output, reducing linear projections to GEMV-class operations with arithmetic intensity~\cite{roofline} proportional to batch size~\cite{pope2023efficient}. Sparse MoE architectures compound this through expert routing: each token activates $K$ experts out of $E$ total, so under uniform routing a serving batch of $B$ tokens yields an effective per-expert batch size of
\begin{equation}
    b_{\text{expert}} = \frac{B \cdot K}{E}
    \label{eq:b_expert}
\end{equation}
For DeepSeek-v3.2 ($K{=}8$, $E{=}256$), a serving batch of $B{=}32$ yields $b_{\text{expert}}{=}1$, i.e., each expert processes a single token, reducing its FFN projections to GEMVs. This pushes each expert's FFN execution into the memory-bandwidth-bound regime: HBM bandwidth is saturated while Tensor Cores remain underutilized. The bottleneck is most acute in interactive decode but persists at moderate serving batch sizes where per-expert token counts remain low.

Profiling per-expert FFN GEMMs with cuBLAS-Lt FP8 on an H100 quantifies how big this memory-bound regime is. The largest per-expert projections (DeepSeek-v3.2 and GLM-5 UpGate) reach 50\% SM compute utilization only at $b_{\text{expert}} \approx 512$, and the smaller Qwen3-30B projection requires $b_{\text{expert}} \approx 4{,}096$ to cross the same point. Inverting (\ref{eq:b_expert}) gives the corresponding serving batches: $B \approx 16{,}384$ tokens for DeepSeek-v3.2 and GLM-5, and $B \approx 65{,}536$ for Qwen3-30B. Realistic interactive-decode batches are orders of magnitude below. So, the common operating regime is squarely in the bandwidth-bound region that FORGE targets.

While Sparse MoE reduces active FLOPs per token, it increases total parameter capacity---often exceeding 600~GB for frontier models. Fitting these parameters in HBM forces multi-node Expert Parallelism (EP)~\cite{gshard, deepspeedmoe, tutel}, where each decode step incurs cross-node all-to-all token routing. This communication overhead scales with EP degree and persists even with high-bandwidth interconnects (NVLink within a node, InfiniBand across nodes). The resulting deployment topology simultaneously consumes HBM capacity that could otherwise serve KV cache scaling and introduces per-token networking latency that compounds across MoE layers.

\textbf{Alternatives?} Quantization (FP8, FP4, INT4)~\cite{fp8formats, awq, gptq} reduces per-parameter storage/transfer cost. It does not change the core access pattern: the GPU must still load every weight from HBM for every token. Our FP8 baseline already uses the narrowest weight format which the H100's Tensor Cores support natively. FORGE is orthogonal to this axis rather than competing with it: it synthesizes the weights of a linear map and is agnostic to the bit resolution at which those weights are stored, so it composes on top of whatever quantization a deployment already uses. The question shifts from how much smaller a weight can be made, to whether it has to be fetched at all.

\subsection{Rotor Projection: Decouple Storage and Compute}
\label{sec:rotor_stages}

{\bf Intuition. } Consider the transformation of a vector via a linear map. The de facto approach is to represent  this weight matrix explicitly.  If we instead view the global transform as a {\em composition} of elementary simpler transformations (e.g., rotations), then we only need to store the specific angles that define them. Pence et al.~\cite{pence2025composing} established that standard linear projections can be replaced and approximated by rotors, structured transformations built from Clifford algebra, achieving competitive downstream performance in LLM attention layers with only $O(\log^2 d)$ learned parameters compared to the $O(d^2)$ of dense matrices. 

\textbf{Rotors and Multivectors.} We know that a rotation matrix acts on vectors. A rotor simply extends this action to the full {\em multivector} space of the algebra, naturally lifting the rotation to operate on the higher-order components. For a more complete introduction, see~\cite{pence2025composing}. The rotor construction replaces a dense $2^n \times 2^n$ weight matrix with a block-diagonal structured matrix, where $d = 2^n$ is the input dimension. For an integer $n$, this matrix decomposes into $n+1$ independent blocks along the diagonal, indexed by $k \in \{0, \ldots, n\}$. The $k$-th block has dimension $\binom{n}{k} \times \binom{n}{k}$, and the block sizes sum to the total dimension, as $\sum_{k=0}^{n} \binom{n}{k} = 2^n$. The blocks at $k=0$ and $k=n$ are trivially the scalar identity. The remaining blocks are dense and must be generated.

\begin{figure}[b]
    \centering
    \includegraphics[width=\linewidth]{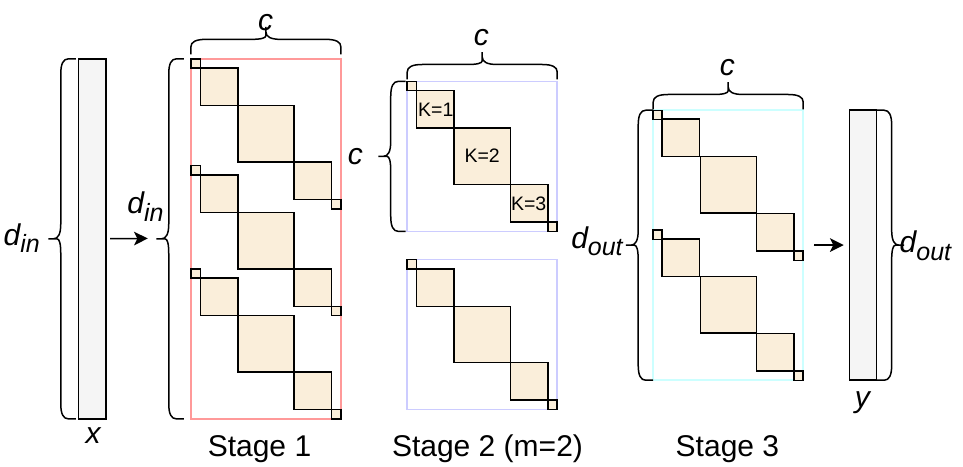}
    \caption{Rotor 3-stage projection pipeline ($n{=}4$, $c{=}16$).}
    \label{fig:rotor_stages}
\end{figure}

\textbf{Building with Blocks.} To handle arbitrary dimensions, the projection is tiled using multiple $2^n \times 2^n$ block-diagonal matrices, which are called \emph{chunks}. Each chunk is an independent instance of the block-diagonal structure with its own learned parameters. A single linear projection then maps to a 3-stage pipeline preserving the original dimensions ($d_{\text{in}} \to d_{\text{out}}$): stage 1 projects from $d_{\text{in}}$ to the chunk size $c = 2^n$ via $\lceil d_{\text{in}}/c \rceil$ chunks, stage 2 applies $m$ rotor maps, each producing a $c \times c$ block-diagonal matrix, and stage 3 projects back from $c$ to $d_{\text{out}}$ via $\lceil d_{\text{out}}/c \rceil$ chunks. We adopt $m=2$ for all projections in this work. The total number of chunks per projection is $\lceil d_{\text{in}}/c \rceil + m + \lceil d_{\text{out}}/c \rceil$, each requiring independent generation.
In this work we set $n = \lfloor \log_2(\min(d_{\text{in}}, d_{\text{out}})) \rfloor$ per projection, yielding $n \in \{9, 10, 11\}$ across the five target MoE models evaluated in Section~\ref{sec:evaluation}. Figure~\ref{fig:rotor_stages} illustrates the resulting 3-stage structure.

Each chunk is parameterized by $\binom{n}{2}$ learned scalar angles, the bivector coefficients in~\cite{pence2025composing}. A rearrangement of the original construction yields the following generation procedure for each $\binom{n}{k} \times \binom{n}{k}$ block within a chunk. Starting from the identity matrix, perform $\binom{n}{2}$ in-place updates, indexing by $t$. Pass $t$ corresponds to rotation axis $(p,q)$ with $0 \leq p < q \leq n{-}1$, and updates a set of disjoint row pairs $(i, j)$ via
\begin{equation}
\begin{bmatrix} W_i \\ W_j \end{bmatrix}
\leftarrow
\begin{bmatrix} \cos(\theta_t) & -\sigma_t \sin(\theta_t) \\ \sigma_t \sin(\theta_t) & \cos(\theta_t) \end{bmatrix}
\begin{bmatrix} W_i \\ W_j \end{bmatrix}
\label{eq:rotation}
\end{equation}
where $W_i, W_j \in \mathbb{R}^{\binom{n}{k}}$ are the row vectors of the block at indices $i$ and $j$, and $(\cos\theta_t, \sin\theta_t)$ are precomputed from the learned parameters. The row-pair set and the sign $\sigma_t \in \{+1, -1\}$ are determined by $(n, k, p, q)$.

\section{Rotor Generation Architecture}
\label{sec:rotor_generation}

Pence et al.~\cite{pence2025composing} showed the theoretical viability of rotor-structured linear projections, but their study was limited to attention layers. We redirect this construction to Sparse MoE FFN decode, where the capacity and bandwidth bottlenecks are most acute. As MoE parameters dominate the total model footprint, the small per-expert batch sizes reduce arithmetic intensity to memory-bound levels. To make procedural generation viable at this scale, their construction must be restructured to expose parallelism for fast, on-the-fly execution during autoregressive decoding.

\textbf{Challenge.} Even after we bound the theoretical workload by adopting $m{=}2$ for all rotor projections (Section~\ref{sec:quality_validation} evaluates its effect on accuracy), we find that systems performance is still bottlenecked. Generating these parallelized weights on-the-fly via standard pipelines ends up starving the Tensor Cores and also creates difficulties for the GPU's instruction scheduler. To adjust the theory to a performant systems architecture, this section derives the parallel execution model from the initial inherently sequential generation concept (Section~\ref{sec:rotor_stages}). Then, we discuss the hardware mandate it implies: why a software-only generation approach hurts the Tensor Cores it is meant to supply. Algorithm~\ref{alg:rotor_gen} states the execution model this section arrives at, and Figure~\ref{fig:worked_example} works one iteration of it through a small block.

\textbf{Key Architectural Insight:} While the mathematical rotor formulation is inherently sequential in its outer loop, its inner combinatorial updates exhibit strict, static data independence. Our fundamental insight is that this static independence can be mapped to a dedicated, collision-free hardware scheduler. This allows us to bypass the massive dynamic scheduling overhead that makes a software-only (SIMT) approach not viable, enabling on-chip weight generation that perfectly overlaps with Tensor Core execution.

\subsection{Extracting Hardware-Friendly Parallelism from Algebraic Structure}
\label{sec:parallel_gen}

\emph{Problem:} The baseline rotor update (Eq.~\ref{eq:rotation}) is formulated as a sequential loop over rotation axes, which is fundamentally incompatible with high-throughput hardware generation. \emph{Solution:} We expose three orthogonal axes of parallelism by decomposing the mathematical structure, mapping it directly to hardware execution units. All three lie inside a single map, not across the $m$ maps where~\cite{pence2025composing} discusses parallelism.

We develop a parallel execution model for the procedure of Section~\ref{sec:rotor_stages} by observing that the row-vector update of Eq.~\ref{eq:rotation} decomposes into independent scalar operations along the column dimension. For any column index $c \in \left[0, \binom{n}{k}\right)$, the update to the scalar pair $(W_{i,c}, W_{j,c})$ depends only on $(W_{i,c}, W_{j,c})$ and the shared parameters $(\cos\theta_t, \sigma_t\sin\theta_t)$. While the outer loop over rotation axes $t$ remains sequential, this per-column decomposition---combined with the structural independence of disjoint row pairs within each pass---yields three orthogonal axes of parallelism that the RGU exploits: (1)~\emph{column independence} within a row-pair update, (2)~\emph{row-pair independence} within a rotation axis, and (3)~\emph{$k$-block independence} within a chunk. We develop each axis below, then show how they combine into the unit of work the RGU generates per request: the \emph{Supertile}.

\paragraph{Column independence.}
The per-column decomposition maps onto a SIMD datapath of width $W_{col}{=}64$, matching standard GPU vector widths and tiled-GEMM tile dimensions.

\paragraph{Row-pair independence.}
Within iteration $t$, the rotation axis $(p,q)$ induces $\binom{n{-}2}{k{-}1}$ independent row pairs: each pair $(i,j)$ is determined by a distinct $(k{-}1)$-element subset of the $n{-}2$ invariant positions (those other than $p$ and $q$), so row indices across pairs are disjoint by construction. These pairs carry no intra-iteration data dependencies---only the boundary between consecutive iterations $t$ and $t{+}1$ is sequential, as the same rows may reappear under a new axis. Because all pair indices are statically determined by the combinatorial structure of $(n, k, t)$, the pair processing order is a free parameter---a property the microarchitecture exploits for conflict-free multi-pair throughput (Section~\ref{sec:sram_pe}).

\begin{figure}[t]
    \centering
    \includegraphics[width=0.85\linewidth]{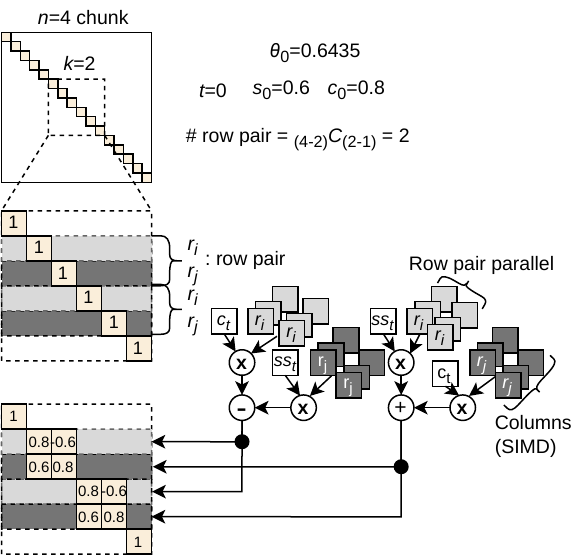}
    \caption{$(n{=}4, k{=}2)$ block, iteration $t{=}0$. Light/dark shading marks the two independent row pairs induced by axis $(p,q)$.}
    \label{fig:worked_example}
\end{figure}

\begin{figure}[b]
    \centering
    \includegraphics[width=0.925\linewidth]{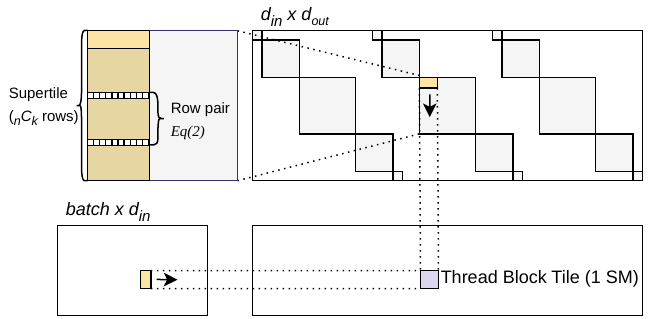}
    \caption{Mapping an on-chip generated Supertile to a standard GEMM Thread Block Tile.}
    \label{fig:rotor_gemm_supertile}
\end{figure}

\paragraph{$k$-block independence and the Supertile.}
The $k$-block independence (each of the $n{+}1$ blocks within a chunk is generated independently) and the column independence together define a natural generation unit: the \emph{Supertile} (Figure~\ref{fig:rotor_gemm_supertile}). A Supertile spans the full $\binom{n}{k}$ rows of one $k$-block and a column width matching the GEMM thread-block tile (64 elements in our configuration)---it is the intersection of the algebraic block-diagonal structure and the tiled-GEMM tile granularity. This alignment means each SM consumes one weight tile per K-dimension iteration, and a single Supertile provides exactly one such tile. Because $\binom{n}{k}$ rows typically exceed the GEMM tile's K-dimension (BlockItemsK~\cite{cutlass}, typically 128 for FP8), the same Supertile feeds $n_{\text{stile}} = \lceil \binom{n}{k} / \text{BlockItemsK} \rceil$ consecutive GEMM iterations along the K-dimension without regeneration. This Supertile reuse factor amortizes the generation latency: the effective per-iteration cost is $T_{\text{supertile\_gen}} / n_{\text{stile}}$.
Figure~\ref{fig:worked_example} makes the row-pair parallelism concrete for an $(n{=}4, k{=}2)$ block. At iteration $t{=}0$, the axis $(p,q)$ induces $\binom{n{-}2}{k{-}1} = \binom{2}{1} = 2$ independent row pairs (light and dark shading), each updated via Eq.~(\ref{eq:rotation}) with shared parameters $(c_t, \sigma_t s_t){=}(0.8, 0.6)$ applied across every column of the block. The two pairs share no source or destination rows, so they execute concurrently without synchronization---the pair-parallel axis the RGU exploits.

Algorithm~\ref{alg:rotor_gen} formalizes the parallel execution model. The outer loop over $t$ is sequential; within each iteration, column updates and row-pair updates execute in parallel.

\begin{algorithm}[t]
\caption{Parallel Rotor Generation for {\em single} $(n,k)$ Block}
\label{alg:rotor_gen}
{\small
\begin{algorithmic}[1]
\Require Parameters $\{(\cos\theta_t, \sin\theta_t)\}_{t=1}^{\binom{n}{2}}$, index maps $\mathcal{I}_t, \mathcal{J}_t$, signs $\sigma_t$
\Ensure $W \in \mathbb{R}^{\binom{n}{k} \times \binom{n}{k}}$
\State $W \leftarrow I_{\binom{n}{k}}$ \Comment{Identity initialization}
\For{$t = 1$ \textbf{to} $\binom{n}{2}$} \Comment{\textit{Sequential}: rotation axes}
    \ForAll{columns $c$ \textbf{in} $[0, \binom{n}{k})$} \Comment{\textit{SIMD-parallel}: 64-wide}
        \ForAll{row pairs $l$ \textbf{in} $[0, \binom{n{-}2}{k{-}1})$} \Comment{\textit{Pair-parallel}: scheduled}
            \State $(i, j) \leftarrow (\mathcal{I}_t[l],\; \mathcal{J}_t[l])$
            \State Apply Eq.~(\ref{eq:rotation}) to $(W_{i,c}, W_{j,c})$
        \EndFor
    \EndFor
\EndFor
\end{algorithmic}
}
\end{algorithm}

\subsection{Hardware Mandate and SM Integration}
\label{sec:hardware_mandate}

\textbf{The SIMT Bottleneck:} A natural first approach is to implement this parallel generation algorithm as a software GPU kernel. We built such a kernel, exploiting the same three axes of parallelism the RGU exploits. Table~\ref{tab:simt_overhead} reports it against the baseline expert GEMM, which includes both the HBM weight load and the GEMM itself. Generation alone takes $25.8\times$ to $45.1\times$ longer than that entire baseline operation. Because the comparison places generation time against the baseline total, it already credits the software path with perfect overlap between generation and the GEMM that consumes it, so no scheduling policy can close this gap.

Two mechanisms are mainly responsible. Instruction issue saturates: the SIMT kernel executes $76\times$ to $149\times$ more instructions than the baseline to deliver the same tiles, instruction-issue rate rather than arithmetic throughput bounds the generation. Shared-memory access is irregular: every update of Eq.~(\ref{eq:rotation}) reads and writes a pair of rows per column, a shape that does not align with the shared-memory bank structure, and bank conflicts rise by two to five orders of magnitude. Across the five models, warp-cycle attribution assigns $49\%$ to $64\%$ of generation time to these two causes together, against $9\%$ to $14\%$ for the baseline.

Neither of the two mechanisms is a tuning gap. A SIMT lane must issue every update of Eq.~(\ref{eq:rotation}) explicitly, so a hundred-fold instruction count is a property of expressing this algorithm on SIMT rather than of the kernel that expresses it, and the paired-row access shape is fixed by Eq.~(\ref{eq:rotation}) while shared memory's banking is fixed by the hardware. \emph{Crucially, this starves the Tensor Cores it is meant to feed;} co-residency with \texttt{wgmma}~\cite{h100wp} instructions destroys pipeline overlap, degenerating execution time toward $T \approx T_{compute} + T_{gen}$. The slowdowns in Table~\ref{tab:simt_overhead} are therefore a lower bound on what a software path actually costs.

\begin{table}[t]
    \centering
    \caption{SIMT weight generation against the baseline expert GEMM.}
    \label{tab:simt_overhead}
    \small
    \begin{tabular}{l r r r}
        \hline
        \textbf{Model} & \textbf{Slowdown} & \textbf{Bank conflicts} & \textbf{Instr.} \\
         & & \textbf{(gen / base)} & \\
        \hline
        DeepSeek-v3.2 & 38.8$\times$ & 1{,}314{,}929 / 15{,}165 & 148.7$\times$ \\
        GLM-5 & 33.8$\times$ & 1{,}026{,}936 / 10{,}981 & 147.5$\times$ \\
        Qwen3-235B & 45.1$\times$ & 868{,}634 / 2{,}919 & 116.9$\times$ \\
        Qwen3-30B & 33.7$\times$ & 644{,}941 / 5 & 76.3$\times$ \\
        Qwen3-Coder-Next & 25.8$\times$ & 124{,}711 / 1 & 80.3$\times$ \\
        \hline
    \end{tabular}
\end{table}

\begin{figure}[!t]
    \centering
    \includegraphics[width=.9\linewidth]{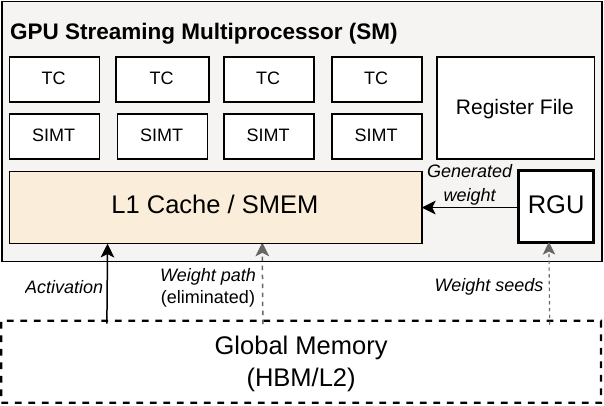}
    \caption{SM integration. The RGU bypasses HBM by generating weight tiles directly into SMEM.}
    \label{fig:sm_integration}
\end{figure}

\textbf{The RGU Solution:} To eliminate this contention, we decouple the generation workload from the SIMT pipeline entirely. We introduce the Rotor Generation Unit (RGU), a compact, fixed-function accelerator embedded within each Streaming Multiprocessor. By utilizing a dedicated pipelined FMA datapath and a local SRAM with a precomputed conflict-free schedule, the RGU bypasses dynamic scheduling bottlenecks and requires no shared SIMT resources. Upon receiving a command with the target $(n, k)$ configuration and rotor seeds, the RGU generates a complete weight Supertile directly into SMEM. This on-chip generation replaces the conventional HBM weight-loading path while remaining fully transparent to the Tensor Cores (Figure~\ref{fig:sm_integration}).

\subsection{Rotor Tiled-GEMM Pipeline}
\label{sec:tiled_gemm}

The defining system-level integration insight of FORGE is dimensional alignment: the algebraic ``$k$-block'' inherently matches the standard GEMM thread-block tile granularity (the Supertile). This alignment dictates our pipeline design.

\begin{figure}[b]
    \centering
    \begin{subfigure}[b]{\linewidth}
        \centering
        \includegraphics[width=0.925\linewidth]{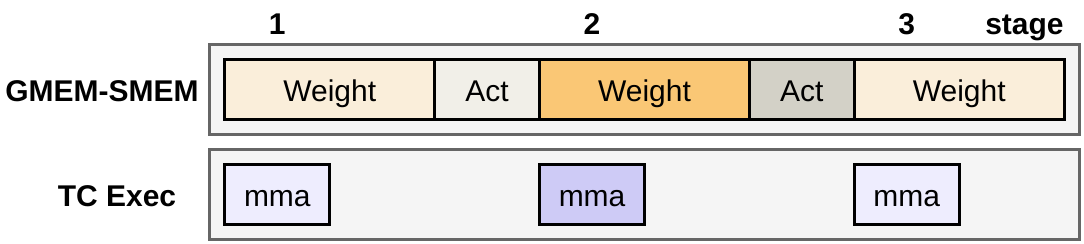}
        \subcaption{Baseline}
        \label{fig:baseline_timing}
    \end{subfigure}
    \begin{subfigure}[b]{\linewidth}
        \centering
        \includegraphics[width=0.925\linewidth]{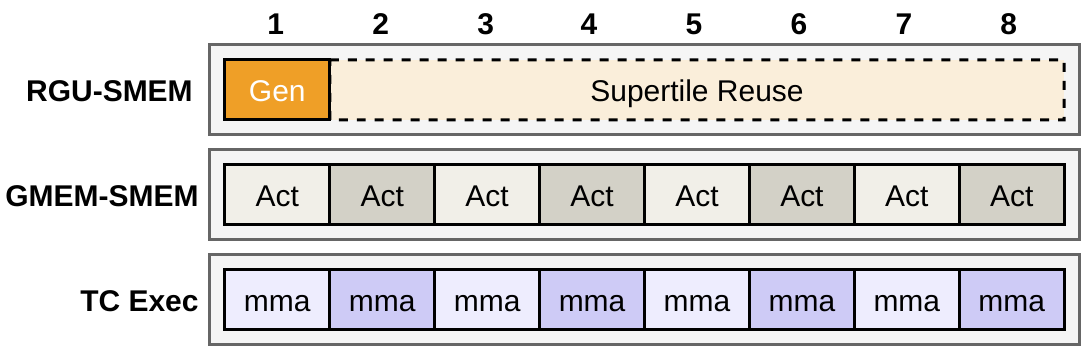}
        \subcaption{FORGE}
        \label{fig:rotor_timing}
    \end{subfigure}
    \caption{Baseline vs. FORGE Tiled-GEMM pipelines.}
    \label{fig:pipeline_timing}
\end{figure}

As shown in Figure~\ref{fig:pipeline_timing}a, the baseline pipeline shares HBM bandwidth between weights and activations, compounding per-iteration latency:
\begin{equation}
    T_{baseline} = \max\!\left(T_{compute},\; T_{mem\_input} + T_{mem\_weight}\right)
\end{equation}

RGU replaces HBM weight-loading with on-chip Supertile generation via a dedicated datapath. This decouples weight delivery from the memory subsystem, allowing generation, activation streaming, and Tensor Core execution to overlap fully (Figure~\ref{fig:pipeline_timing}b):
\begin{equation}
    T_{rotor} = \max\!\left(T_{compute},\; T_{mem\_input},\; \frac{T_{\text{supertile\_gen}}}{n_{\text{stile}}}\right)
    \label{eq:rotor_time}
\end{equation}
When amortized generation latency falls below the HBM transfer time it replaces, execution transitions from bandwidth-bound to compute-bound.

Within each expert, the RGU generates Supertiles on-demand following the standard SwiGLU FFN dataflow (Up/Gate $\rightarrow$ Down). The total generation workload per expert sums the chunks across all three projections, each independently requiring $m=2$ Stage-2 Rotor maps (Section~\ref{sec:rotor_stages}).

\section{RGU Microarchitecture}
\label{sec:microarch}

We recall notation from Section~\ref{sec:rotor_stages} and Section~\ref{sec:parallel_gen} (Equation~\ref{eq:rotation}, Algorithm~\ref{alg:rotor_gen}). A chunk is one $2^n \times 2^n$ block-diagonal matrix ($n = 9{\ldots}11$ for MoE workloads), which the RGU materializes one Supertile at a time: $n{+}1$ blocks indexed by $k$, each $\binom{n}{k} \times \binom{n}{k}$. Generation proceeds through $\binom{n}{2}$ iterations indexed by $t$. Iteration $t$ selects rotation axis $(p, q)$ ($0 \le p < q \le n{-}1$) and updates $\binom{n{-}2}{k{-}1}$ independent row pairs $(r_i, r_j)$ using learned parameters $\cos\theta_t$ and $\sigma_t\sin\theta_t$, where $\sigma_t \in \{+1, -1\}$ is a combinatorial sign parity. Columns are independent; the datapath processes 64 elements in parallel. We present the 64$\times$2 dual-pair design.

\subsection{Datapath Overview}

\begin{figure*}[t]
    \centering
    \includegraphics[width=0.9\textwidth]{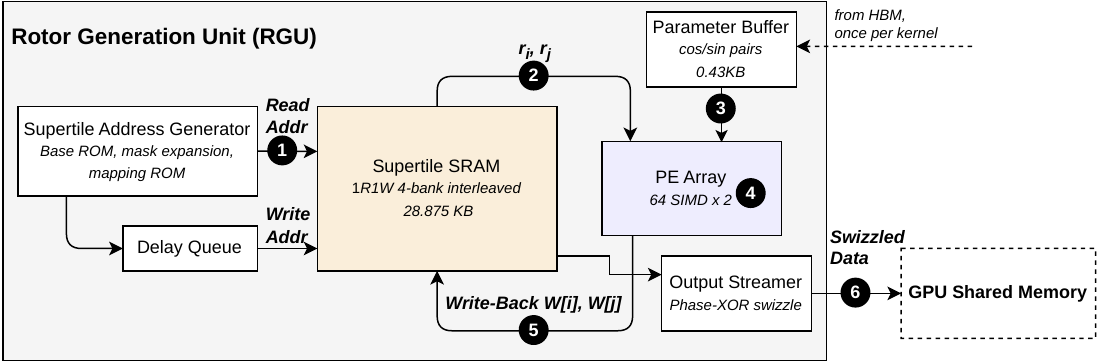}
    \caption{RGU microarchitecture datapath.}
    \label{fig:rgu_microarch}
\end{figure*}

The RGU must generate a full FP8 Supertile---28.875~KB covering up to $\binom{11}{5}{=}462$ rows $\times$ 64 elements from a compact 0.43~KB parameters, completing before the Tensor Core pipeline demands the next weight tile (Section~\ref{sec:tiled_gemm}). The datapath addresses this throughput constraint through a tightly interleaved pipeline (Fig.~\ref{fig:rgu_microarch}): the Supertile Address Generator (SAG) synthesizes physical row-pair addresses from compact ROMs (\ding{182}; Section~\ref{sec:sag}), and operand rows $r_i$, $r_j$ are read from a 4-bank interleaved Supertile SRAM (\ding{183}) that gives 2R2W effective bandwidth from simple 1R1W banks. Stationary $(\cos\theta_t, \sin\theta_t)$ parameters broadcast from the parameter buffer (\ding{184}) to dual 64-wide PE lanes (\ding{185}), which process two independent row pairs per 2-cycle slot. Updated rows $W[i]$, $W[j]$ write back to SRAM (\ding{186}), with a delay queue synchronizing write-back addresses with the pipelined FMA latency, yielding 100\% PE utilization (under per-bank 1R1W port constraint).

A compact 1R1W parameter buffer stores $\binom{n_{\max}}{2} \times 4 = 220$ precomputed $(\cos\theta_t, \sin\theta_t)$ pairs (one per rotation axis, replicated across up to 4 vertically stacked chunks for multi-chunk projections in Stages~1 and~3), totaling $220 \times 2 = 440$~B (${\approx}0.43$~KB). Storing all chunks' parameters avoids HBM re-fetching between consecutive Supertile generations. Parameters are latched once at each axis boundary $t$ and broadcast to the PE array across the hundreds of cycles in which $\binom{n{-}2}{k{-}1}$ row pairs are processed, eliminating redundant SRAM reads.

Upon generation, the Output Streamer transfers the completed Supertile into GPU Shared Memory (\ding{187}), applying the Phase-XOR swizzle required by Tensor Core instructions directly during write-out. Unaligned $\binom{n}{k}$ heights are zero-padded to the nearest multiple of 16. For trivial identity blocks ($k{=}0$ and $k{=}n$), the streamer bypasses generation entirely and injects explicit \texttt{1.0} scalars directly into SMEM, incurring zero generation cycles.

\subsection{Supertile Address Generator}
\label{sec:sag}

\emph{Problem:} Generating the $(i, j)$ row-pair traces dynamically requires complex control logic, but explicitly storing the trace for an $n{=}11$ configuration requires tens of thousands of entries---far exceeding practical on-chip SRAM limits. \emph{Solution:} The SAG exploits the symmetric algebraic properties of the rotor groups to dynamically synthesize physical addresses from a highly compressed state.

A naive scheme would store the full $(i,j)$ row-pair trace for each $(n,k)$ configuration---tens of thousands of entries for $n{=}11$ alone. The SAG eliminates this by exploiting the algebraic structure: all $\binom{n{-}2}{k{-}1}$ independent row pairs within a single rotation axis $(p,q)$ share a common $(n{-}2)$-bit base combination pattern. A 0.5~KB Base ROM stores all $2^{n_{\max}{-}2}{=}512$ possible base patterns, organized by Hamming weight. A combinational expansion network injects the current axis bits $p$ and $q$ into each base pattern to produce the two operand-row masks ($M_i$ with a $1$ at position $p$, $M_j$ with a $1$ at $q$) on consecutive cycles, time-multiplexing both addresses onto a single ROM port and aligning with the 2-cycle FMA cadence under the 1R1W SRAM constraint.

A 9~KB Mapping ROM translates the expanded 11-bit row masks into dense physical SRAM indices $[0, \binom{n}{k}{-}1]$, jointly keyed by an $n$-mode tag so that a single table serves all $n \in \{9,10,11\}$. The sign $\sigma_t$ is computed combinationally from a parity reduction over the base-pattern bits between positions $p$ and $q$, requiring no stored sign tables. A delay queue synchronizes translated write-back addresses with the PE pipeline, ensuring addresses arrive when pipelined FMA results are ready for commit.

\subsection{Supertile SRAM \& Processing Element}
\label{sec:sram_pe}

\emph{Problem:} Sustaining dual-pair processing throughput requires four simultaneous SRAM accesses (2 reads, 2 writes) per cycle slot. Implementing this with multi-ported SRAM would severely bloat the SM area overhead. \emph{Solution:} We achieve 4-port performance using compact 1R1W banks by exploiting the static predictability of the rotor accesses.

Rather than employing expensive multi-ported SRAM, the design uses 4-bank interleaving (depth 116 per bank) to synthesize 2R2W effective bandwidth from four simple 1R1W banks. Total capacity: $462 \times 512\text{-bit} = 28.875$~KB (FP8), sized for the largest $k$-block ($\binom{11}{5}{=}462$ rows). Collision freedom is guaranteed by a precomputed pair schedule (described below), eliminating dynamic bank arbitration entirely.

\begin{figure}[b]
    \centering
    \includegraphics[width=0.925\linewidth]{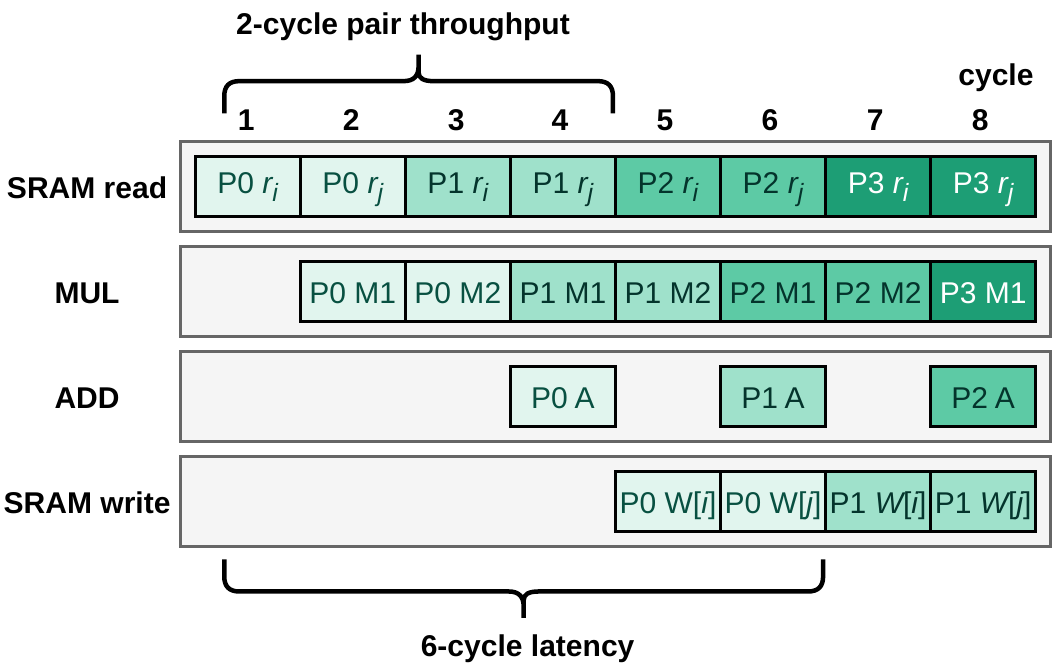}
    \caption{Steady-state PE pipeline timing.}
    \label{fig:pe_timing}
\end{figure}

Each of the two 64-wide SIMD datapaths implements a 3-cycle pipelined FMA split as two multiply stages followed by an add, halving the critical combinational depth. The two multiply stages read $r_i$ and $r_j$ from SRAM on successive cycles and compute the four cross-products required by Eq.~(\ref{eq:rotation}); the add stage commits the rotated row pair $W[i]$, $W[j]$. The add of the \emph{previous} row pair overlaps with the first multiply of the \emph{current} pair, sustaining 2-cycle pair throughput from the 3-cycle pipeline (Figure~\ref{fig:pe_timing}).

\textbf{Static Collision-Free Scheduling:} The dual-pair datapath requires two independent row-pair indices per 2-cycle slot, which would otherwise demand dynamic bank arbitration with cycle-level conflict detection. Instead, a precomputed \emph{Schedule ROM} stores collision-free pair orderings for all $(n,k)$ configurations with $n{=}9{\ldots}11$, generated offline via maximum matching over the bank-conflict graph. A single table serves every configuration, eliminating runtime arbitration entirely.

\subsection{Global Control Logic}

A global FSM sequences through all $\binom{n}{2}$ rotation axes $(p, q)$ in reversed lexicographic order (e.g., for $n{=}11$: $(9,10) \rightarrow (8,10) \rightarrow (8,9) \rightarrow \cdots \rightarrow (0,1)$). Within each axis, the FSM counts through $\binom{n{-}2}{k{-}1}$ independent row pairs via the $c\_idx$ counter, driving the SAG and Schedule ROM in lockstep. Inter-$k$-block transitions are managed as contiguous horizontal sweeps across the block-diagonal structure. Zero-initializing the Supertile SRAM prior to generation leaves off-diagonal regions structurally correct without an explicit zero-fill pass.

For trivial identity blocks ($k{=}0$ and $k{=}n$), the FSM bypasses PE execution entirely. At rotation axis boundaries---when the FSM advances from axis $t$ to $t{+}1$ with potentially overlapping rows---in-flight writes from axis $t$ may conflict with the first reads of axis $t{+}1$, creating RAW hazards. Rather than a dynamic scoreboard, the FSM injects a static 5-cycle stall bubble at each boundary, guaranteeing all pending writes drain before the next axis begins (Section~\ref{sec:evaluation}).

\section{Evaluation}
\label{sec:evaluation}

\subsection{Methodology}
\label{sec:methodology}

Baseline per-expert MoE FFN execution times are obtained from NVIDIA Nsight Compute (NCU) profiling of cuBLASLt FP8 GEMMs on an H100 SXM5 GPU (132 SMs, 3.35~TB/s HBM3, CUDA 13.2). For each model, we profile the three expert projections (Up, Gate, Down) at their native dimensions and extract per-tile compute time, memory transfer time, and achieved bandwidth utilization. Rotor execution times are derived from the RTL-verified cycle model of the RGU (Section~\ref{sec:microarch}), converting generation cycles to wall-clock time at the achieved synthesis frequency. We do not evaluate a narrower weight format: FP8 is the narrowest the H100's Tensor Cores support natively (Section~\ref{sec:inference_bottleneck}).

We evaluate five Sparse MoE architectures spanning three chunk dimensions: DeepSeek-v3.2 and GLM-5 ($n{=}11$, 256 experts), Qwen3-235B ($n{=}10$, 128 experts), and Qwen3-30B and Qwen3-Coder-Next ($n{=}9$, 128/512 experts). Kernel-level evaluation targets the memory-bound decode regime; system-level analysis (Section~\ref{sec:system_impact}) extends to higher serving batch sizes where freed HBM capacity drives throughput gains independent of per-kernel speedup. We report all system results at $r{=}0.3$, chosen as a conservative point on the accuracy-performance tradeoff (Section~\ref{sec:quality_validation}).

The RGU is implemented in SystemVerilog as an FP8 dual-pair architecture processing two independent row pairs per 2-cycle slot. Standard cells are synthesized using Synopsys Design Compiler (DC X-2025.06) with the ASAP7~\cite{asap7} predictive 7nm finFET library (RVT, TT corner, 0.7\,V/25\textdegree C). SRAM area, energy, and leakage are modeled with CACTI~\cite{cacti} at 32\,nm and scaled to 7\,nm via the polynomial factors of Stillmaker and Baas~\cite{stillmaker2017}. Cross-check (45\,nm) agrees within 11\%.

Full-model decode latency and system throughput are evaluated using the LIMINAL~\cite{liminal} analytical decode simulator. LIMINAL is an externally published framework validated against real-hardware LLM execution to within $7.6\%$ mean absolute error~\cite{liminal}; it composes independently modeled components---Attention weight loading, KV cache access, MoE FFN execution, TP allreduce, EP all-to-all, and per-MoE-layer kernel launch overhead (symmetrically applied to both baseline and rotor paths). We reproduce its per-primitive cost formulas unchanged and supply only the per-expert MoE FFN cost, itself anchored in measured inputs: NCU-profiled cuBLASLt FP8 GEMMs for the baseline path and the RTL-verified RGU cycle model for the rotor path. The baseline topology uses TP+EP with the minimum GPU count to hold model weights in HBM under a $0.9{\times}$ safety margin, leaving headroom for KV cache and activation tensors on the baseline deployment; the FORGE topology replicates seeds across GPUs rather than sharding them. Three context lengths are evaluated: $T \in \{4096, 32768, 131072\}$.

\begin{table}[h]
    \centering
    \caption{System Configuration and Analyzed Workloads. Details baseline GPU parameters (H100), Target LLMs, and RTL synthesis environments.}
    \label{tab:methodology}
    \small
    \begin{tabular}{lp{5.2cm}}
        \hline
        \textbf{Component} & \textbf{Specification} \\
        \hline
        Baseline GPU & H100 SXM5 (132 SMs, 3.35 TB/s HBM3) \\
        Kernel Profiling & cuBLASLt FP8, CUDA 13.2, NCU \\
        Target Models & DeepSeek-v3.2, GLM-5, Qwen3-235B, Qwen3-30B, Qwen3-Coder-Next \\
        Logic Synthesis & Synopsys DC + ASAP7~\cite{asap7} (7nm finFET, RVT, TT) \\
        SRAM Modeling & CACTI~\cite{cacti} 32\,nm scaled to 7\,nm via~\cite{stillmaker2017} \\
        System-Level Sim & LIMINAL~\cite{liminal} analytical per-layer compose (Attn + KV + MoE + Comm + kernel launch) \\
        \hline
    \end{tabular}
\end{table}

\subsection{Organization of Results}
We systematically examine the following questions in our empirical evaluation.

\begin{itemize}
\item First we look at what speedup is possible compared to ``native MoE'' at the operator/kernel simply considering the MoE layer in isolation (Section~\ref{subsec:kernelspeed}).

\item Second, we look at end-to-end performance: both decode latency per user (user tok/sec) and full system performance (system tok/sec) (Section~\ref{sec:decode_latency}).

\item We then look at the detailed hardware implementation efficiency of RGU by examining its RTL implementation (Section~\ref{sec:ppa}).

\item Finally, we look at the ``model-level'' impact of FORGE, measuring perplexity and downstream accuracy on the models and benchmarks introduced above (Section~\ref{sec:quality_validation}).
\end{itemize}

\subsection{Kernel-Level MoE FFN Speedup}\label{subsec:kernelspeed}

\noindent\textit{\textbf{Highlight.} Replacing the HBM weight-load path with on-chip generation, the measured FP8 baseline and the RTL-verified generation latency give a $2.3$--$3.2\times$ per-expert kernel speedup at 8 tokens/expert, marking the transition from memory-bound to compute-bound execution on replaced layers.}
\medskip

Table~\ref{tab:kernel_speedup} presents the per-expert MoE FFN kernel speedup for all five target models under the $m{=}2$ configuration at two operating points: 8 and 16 routed tokens per expert. At 8 tokens/expert, the speedup is $2.3\times$ to $3.2\times$ (FP8), combining the measured baseline with the RTL-verified generation latency through Eq.~(\ref{eq:rotor_time}). Speedup peaks at $n{=}10$ (Qwen3-235B, $3.23\times$), where the ratio of HBM access time to generation time is most favorable. At $n{=}11$ (DeepSeek-v3.2, GLM-5), the $\binom{n}{2}{=}55$ rotation steps increase generation time, yielding $2.3$--$2.5\times$. At $n{=}9$ (Qwen3-30B, Qwen3-Coder-Next), the baseline matrices are small enough that fixed cuBLASLt launch overhead limits the memory-bound advantage, yielding $2.3\times$. At 16 tokens/expert, speedups narrow to $1.13\times$--$1.67\times$ as compute time dominates, confirming that the RGU targets the memory-bound regime where per-expert token counts are small. Even at these moderate per-kernel speedups, the capacity-driven throughput gains remain substantial (Section~\ref{sec:system_impact}).

In absolute terms, the RGU reduces per-expert MoE FFN latency from $23.48$~$\mu s$ to $10.06$~$\mu s$ for DeepSeek-v3.2 and from $14.58$~$\mu s$ to $4.52$~$\mu s$ for Qwen3-235B (FP8, 8 tokens/expert).

\begin{table}[!b]
    \centering
    \caption{Per-MoE-FFN Kernel Speedup (FP8, RGU, $m{=}2$). Baseline: cuBLASLt FP8 GEMM on H100.}
    \label{tab:kernel_speedup}
    \small
    \begin{tabular}{l c r r}
        \hline
        \textbf{Model} & $n$ & \textbf{8 tok/expert} & \textbf{16 tok/expert} \\
        \hline
        DeepSeek-v3.2 & 11 & 2.33$\times$ & 1.29$\times$ \\
        GLM-5 & 11 & 2.47$\times$ & 1.40$\times$ \\
        Qwen3-235B & 10 & 3.23$\times$ & 1.67$\times$ \\
        Qwen3-30B & 9 & 2.28$\times$ & 1.13$\times$ \\
        Qwen3-Coder-Next & 9 & 2.34$\times$ & 1.19$\times$ \\
        \hline
    \end{tabular}
\end{table}

\subsection{Full-Model Decode Latency}
\label{sec:decode_latency}

\noindent\textit{\textbf{Highlight.} With $30\%$ of the MoE layers replaced, FORGE is projected to reduce end-to-end single-token decode latency by $1.10$--$1.15\times$ at short context ($T{=}4096$). The advantage persists at $1.07$--$1.11\times$ through $T{=}128$K as KV cache access becomes the dominant cost.}
\medskip

Table~\ref{tab:decode_speedup} presents the full-model single-token decode latency speedup across five models and three context lengths, with rotors configured at the conservative $r{=}0.3$ partial-replacement point evaluated for accuracy in Section~\ref{sec:quality_validation}. At $T{=}4096$, the projected decode speedup runs from $1.10\times$ (DeepSeek-v3.2) to $1.15\times$ (Qwen3-30B). The gain comes from faster MoE FFN execution on the rotor-replaced layers ($2.3$--$3.2\times$ per-expert kernel speedup, Tab.~\ref{tab:kernel_speedup}); the kept baseline MoE layers still pay HBM weight loads and EP all-to-all, charged separately in the per-layer decode model. At longer context ($T{=}131072$), speedups narrow to $1.07$--$1.11\times$ because KV cache access time grows linearly with context length and is identical across configurations, diluting the MoE FFN contribution. This is inherent to any MoE-FFN-focused optimization and does not diminish the capacity and throughput advantages analyzed in Section~\ref{sec:system_impact}.

\begin{figure}[t]
    \centering
    \includegraphics[width=0.875\columnwidth]{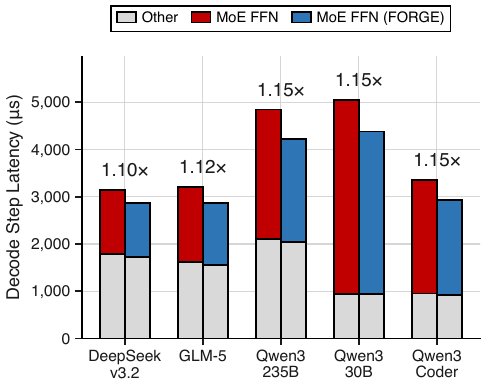}
    \caption{Full-model decode latency (FP8, B=1, T=4096, r=0.3).}
    \label{fig:decode_comparison}
\end{figure}

Fig.~\ref{fig:decode_comparison} shows the per-layer time decomposition at $T{=}4096$. For Qwen3-30B, MoE FFN dominates baseline per-layer time on a single GPU, so even at $r{=}0.3$ the $2.28\times$ kernel speedup on the replaced layers translates into $1.15\times$ full-model decode reduction. For DeepSeek-v3.2, the MoE FFN fraction is smaller due to larger Attention weights and per-token EP all-to-all cost, yielding $1.10\times$ decode. The replacement-ratio ablation (Fig.~\ref{fig:replace_ablation}) quantifies how this gap closes as $r$ grows.

At $r{=}0.3$, the minimum-fit GPU count for DeepSeek-v3.2 and GLM-5 drops from $10$/$11$ to $7$/$8$, fitting both models in a single NVLink-connected node and removing cross-node all-to-all from the decode path. Under a hypothetical $r{=}1.0$ extrapolation---reported only as a design-knob upper bound---both models collapse to a single device.
\begin{table}[!b]
    \centering
    \caption{Full-Model Decode Latency Speedup (FP8, RGU, $B{=}1$, $r{=}0.3$). Baseline: TP+EP with minimum GPUs to fit weights. FORGE replaces $30\%$ of the MoE layers; kept baseline MoE layers still pay EP all-to-all.}
    \label{tab:decode_speedup}
    \small
    \begin{tabular}{l r r r}
        \hline
        \textbf{Model} & $T{=}4\text{K}$ & $T{=}32\text{K}$ & $T{=}128\text{K}$ \\
        \hline
        DeepSeek-v3.2 & 1.10$\times$ & 1.09$\times$ & 1.07$\times$ \\
        GLM-5 & 1.12$\times$ & 1.11$\times$ & 1.08$\times$ \\
        Qwen3-235B & 1.15$\times$ & 1.12$\times$ & 1.08$\times$ \\
        Qwen3-30B & 1.15$\times$ & 1.14$\times$ & 1.11$\times$ \\
        Qwen3-Coder-Next & 1.15$\times$ & 1.13$\times$ & 1.09$\times$ \\
        \hline
    \end{tabular}
\end{table}

\begin{figure*}[t]
    \centering
    \includegraphics[width=\textwidth]{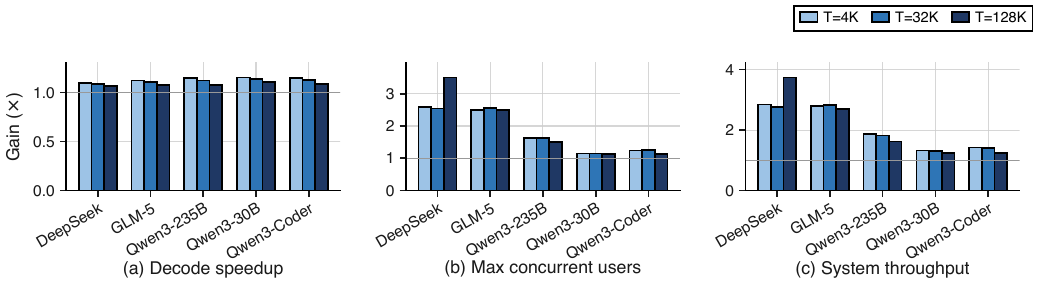}
    \caption{Context-length ablation (matched GPU count). (a)~Decode speedup narrows as KV cache access dilutes the MoE FFN contribution. (b)~Freed HBM grows max concurrent users per GPU sharply at long context. (c)~The user-count gain offsets the narrowing decode speedup, so TPS gains hold or widen.}
    \label{fig:context_throughput_ablation}
\end{figure*}

\subsection{System-Level Throughput and Capacity}
\label{sec:system_impact}

\noindent\textit{\textbf{Highlight.} Freed HBM from the replaced layers expands KV cache capacity by $1.2$--$2.6\times$, which compounds with faster decode to give a projected $1.3$--$2.8\times$ system throughput at matched GPU count, with the largest gains landing on the heaviest MoE models (DeepSeek-v3.2, GLM-5).}
\medskip

System throughput (matched GPU count) compounds two gains:
\begin{equation}
\text{System TPS} = N_{\text{GPUs}} \times U_{\text{max}} \times \frac{1}{t_{\text{decode}}}
\label{eq:system_tps}
\end{equation}
At $r{=}0.3$, FORGE improves both terms. First, faster MoE FFN execution is projected to cut per-token latency $t_{\text{decode}}$ by $1.10$--$1.15\times$ (Table~\ref{tab:decode_speedup}). Second, the freed weight HBM, $28$--$29\%$ of total parameter memory, expands maximum concurrent users $U_{\text{max}}$ (computed as available memory divided by per-user KV size, under a $0.9\times$ safety margin) by $1.2$--$2.6\times$ at $T{=}4096$. Together, these yield the throughput gains shown in Figure~\ref{fig:context_throughput_ablation}c.

At longer contexts ($T{=}131072$), individual users consume more KV memory, making the freed HBM even more impactful. The $U_{\text{max}}$ advantage grows to $1.1$--$3.5\times$, directly countering the per-token latency degradation observed in Table~\ref{tab:decode_speedup} and ensuring throughput gains persist or widen (Figure~\ref{fig:context_throughput_ablation}b).

\begin{figure}[b]
    \centering
    \includegraphics[width=0.925\columnwidth]{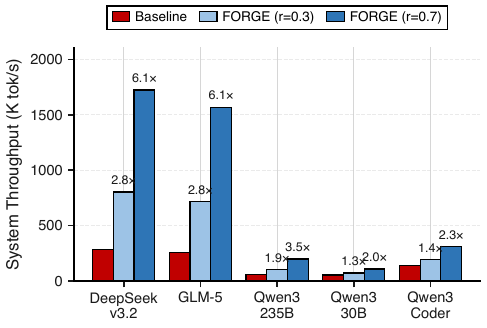}
    \caption{System throughput scaling across replacement ratios (r) (T=4096). Accuracy is evaluated at $r{=}0.3$ only; $r{=}0.7$ is a what-if point (Section~\ref{sec:system_impact}).}
    \label{fig:replace_ablation}
\end{figure}

Finally, FORGE allows graduated substitution (Figure~\ref{fig:replace_ablation}). While $r{=}0.3$ delivers up to $2.8\times$ throughput over the baseline, increasing the replacement fraction to $r{=}0.7$ pushes this to $2.0$--$6.1\times$. Larger models benefit disproportionately, as even partial replacement frees enough HBM to meaningfully expand user capacity while simultaneously cutting the memory-bound weight-load path. We have not evaluated accuracy at $r{=}0.7$, so this point is a what-if: it shows the throughput available to future models or recovery methods that sustain higher replacement without substantial accuracy loss.

\subsection{Microarchitectural PPA}
\label{sec:ppa}

\noindent\textit{\textbf{Highlight.} At ${\sim}4.2$~$mm^2$ ($0.5\%$ of the H100 die) and ${\sim}5.1$~W ($0.7\%$ of its TDP) aggregated across 132 SMs, the RGU is a negligible on-chip overhead that removes hundreds of gigabytes of HBM expert-weight storage.}
\medskip

Table~\ref{tab:ppa_summary} summarizes the 7\,nm PPA for the FP8 dual-pair RGU. Because the datapath relies on compact 4-bit mantissa multipliers (dominating the logic area), the full unit occupies just $0.032$~$mm^2$ per SM. Aggregated across the 132 SMs of an H100 SXM5, the RGU array requires only ${\sim}4.2$~$mm^2$---roughly $0.5\%$ of the 814~$mm^2$ H100 die~\cite{h100wp}---yet entirely removes hundreds of gigabytes of HBM expert-weight storage. ASAP7 is a predictive academic node, while the H100 baseline is built on a commercial 4\,nm process, so these figures are conservative: a commercial process (e.g., TSMC N4/N5) would yield a smaller footprint and a higher clock.

The pipeline closes timing at 1.80~GHz, safely reaching 91\% of the H100's boost clock. Total steady-state power is ${\sim}39$~mW per SM, aggregating to ${\sim}5.1$~W system-wide---a negligible $0.7\%$ of the H100's 700~W TDP. (RTL behavior is fully verified against a reference rotor model across all 27 valid $(n,k)$ configurations.)

\begin{table}[h]
    \centering
    \caption{7 nm PPA Summary for the FP8 dual-pair RGU.}
    \label{tab:ppa_summary}
    \small
    \begin{tabular}{l r l}
        \hline
        \textbf{Component} & \textbf{Value} & \textbf{Notes} \\
        \hline
        Logic Area & 0.017~$mm^2$ & PE array ${\sim}$69\% \\
        SRAM Area & 0.015~$mm^2$ & 28.875~KB, 4-bank \\
        \textbf{Total / SM} & \textbf{0.032~$mm^2$} & \\
        \hline
        vs one SM (3.475~$mm^2$~\cite{locuza_ada}) & ${\sim}$0.9\% & \\
        \hline
        Clock (ASAP7) & 1.80~GHz & WNS $-65$~ps @ 2~GHz \\
        Logic Power (0.7\,V) & 35.9~mW & 3.4 switch + 32.5 internal \\
        SRAM Leakage (7\,nm) & ${\sim}$3~mW & CACTI 32\,nm scaled \\
        \hline
    \end{tabular}
\end{table}

\subsection{Model Quality Validation}
\label{sec:quality_validation}

\noindent\noindent\textit{\textbf{Highlight.} Across three model families, FORGE at $r{=}0.3$ falls short of the same model fine-tuned with LoRA and no rotor replacement. This accuracy cost is what FORGE trades for the system gains of Sections~\ref{sec:decode_latency}--\ref{sec:system_impact}.}

\medskip

FORGE replaces dense expert weights with a structured approximation, so the natural question is whether the resulting model still behaves well on downstream tasks. To this end, we evaluate model quality across three MoE models (ERNIE-21B, Qwen3-30B, Qwen3-Coder-Next) on WikiText perplexity, HellaSwag accuracy, and ARC-C accuracy. For each model, we show three configurations: the original pretrained baseline, the same model fine-tuned with LoRA rank-8 (no rotor replacement), and FORGE at $r{=}0.3$ with the same LoRA setup applied. The results are shown in Table~\ref{tab:quality_bench}. The LoRA-only model is the reference for FORGE, since the two share the same recovery and differ only in rotor replacement.

\textbf{\textit{Evaluation protocol.}} All downstream numbers are produced with the EleutherAI Language Model Evaluation Harness~\citep{eval-harness}. Every model is wrapped in the harness's HuggingFace backend and scored with the harness's built-in task definitions. HellaSwag (full validation split) and ARC-Challenge (full test split) are scored zero-shot as a log-likelihood ranking over the candidate answers, and we report accuracy as the fraction of examples whose correct answer receives the highest log-likelihood. WikiText-2 perplexity uses the raw test split tokenized with each model's native tokenizer, cut into 512-token windows with stride 256, and is the exponential of the mean token negative log-likelihood. Every reported number is a single, deterministic full pass over the evaluation set, so there is no randomness at evaluation time.

FORGE falls short of the LoRA-only model (Table~\ref{tab:quality_bench}). Under an identical rank-8 LoRA budget, the dense model reaches 6.53 perplexity on Qwen3-30B and FORGE reaches 7.65. This is an expected behavior, as rotor replacement is a structured but lossy approximation, so it necessarily consumes part of the quality headroom that LoRA provides. This gap is the explicit quality cost of procedural weight generation, traded for the memory and bandwidth gains. The size of this gap is set by $r$, which a deployer selects against a memory budget rather than a fixed property of the parameterization.

\begin{table}[!t]
\centering
\small
\caption{Model performance with $r{=}0.3$ on \textit{ERNIE-21B}, \textit{Qwen3-30B}, and \textit{Qwen3-Coder-Next} on WikiText, ARC-C, and HellaSwag.  Baseline + LoRA is the reference for FORGE. The gap between the two is the accuracy cost of rotor replacement.}
\label{tab:quality_bench}
\begin{tabular}{@{}lrrr@{}}
\toprule
Config & WikiText$\downarrow$ & ARC-C$\uparrow$ & HellaSwag$\uparrow$ \\
\midrule
\multicolumn{4}{@{}l}{\textit{ERNIE-21B}} \\
\quad Baseline        & 8.37 & 58.00 & 61.10 \\
\quad Baseline + LoRA & 5.93 & 58.96 & 62.50 \\
\quad FORGE           & 7.16 & 57.25 & 59.30 \\
\midrule
\multicolumn{4}{@{}l}{\textit{Qwen3-30B}} \\
\quad Baseline        & 9.40 & 52.60 & 59.50 \\
\quad Baseline + LoRA & 6.53 & 55.72 & 60.88 \\
\quad FORGE           & 7.65 & 51.36 & 57.10 \\
\midrule
\multicolumn{4}{@{}l}{\textit{Qwen3-Coder-Next}} \\
\quad Baseline        & 8.55 & 60.32 & 64.06 \\
\quad Baseline + LoRA & 5.99 & 61.86 & 63.99 \\
\quad FORGE           & 6.88 & 60.41 & 60.52 \\
\bottomrule
\end{tabular}
\end{table}

\subsection{Limitations}
\label{sec:limitations}

Our baseline is measured and our design is simulated, so the two sides of every speedup rest on different evidence. The numbers fall into three classes. The FP8 baselines come from cuBLASLt on an H100. The RGU numbers come from its RTL. Generation latency comes from a cycle model verified bit-exact against a golden reference, and area, power, and frequency come from synthesis. The kernel speedups in Table~\ref{tab:kernel_speedup} combine the first two through Equation~(\ref{eq:rotor_time}). Everything downstream passes through the analytical decode simulator of Section~\ref{sec:methodology}: the decode latencies in Table~\ref{tab:decode_speedup} and Figure~\ref{fig:decode_comparison}, and the capacity and throughput results in Figures~\ref{fig:context_throughput_ablation} and~\ref{fig:replace_ablation}. It resolves per-primitive cost and does not capture contention or occupancy effects inside a modified SM.

Our synthesis covers the RGU alone. The reported area and power are the unit's and exclude the cost of wiring it into an SM. They also come from a predictive 7\,nm node, while the H100 baseline is built on a commercial 4\,nm process (Section~\ref{sec:ppa}). We expect the integration cost to be small. FORGE replaces the weight-load path and leaves the rest of the pipeline untouched. The Tensor Core consumes activation and weight tiles exactly as before, and the ISA gains one instruction family parameterized by tile shape. Neither change touches the SIMD execution model.

\section{Related Work}
\label{sec:related_work}

\noindent\textbf{Weight Compression.} Post-training quantization~\cite{awq,gptq,smoothquant,llmint8,fp8formats,qserve,marlin,quip,bitnet} and pruning~\cite{sparsegpt,wanda,deepcompression,mishra2021sparsity} shrink the per-weight payload but still fetch every activated weight from HBM. QuaRot~\cite{quarot} and SpinQuant~\cite{spinquant} rotate weights to flatten outliers before quantization, but the rotated tensor is still materialized and stored. Ecco~\cite{ecco} applies entropy-aware compression to cache-resident data, reducing what crosses the memory hierarchy rather than removing the fetch. We \emph{synthesize} the tile on demand from a compact seed, removing the HBM transfer.

\noindent\textbf{Structured Parameterization.} Low-rank and structured factorizations such as LoRA~\cite{lora,qlora}, Monarch~\cite{monarch}, butterfly~\cite{butterfly}, HashedNets~\cite{hashednets}, and Fastfood~\cite{fastfood} all reduce the effective parameter count, and HyperNetworks~\cite{hypernetworks} generate a larger network's weights from a smaller one. While these methods reduce parameters, their compressed representations remain too large for on-chip storage and store the compressed factors in HBM before use. The $O(\log^2 d)$ parametrization of rotors approximates each $O(d^2)$ projection ~\cite{pence2025composing} and is small enough for the regeneration of the tile directly into on-chip SMEM, bypassing HBM.

\noindent\textbf{Sparse MoE Systems and Scheduling.} Top-$k$ routing~\cite{shazeer2017moe,switchtransformer,gshard} drives frontier MoE designs~\cite{mixtral,deepseekmoe,deepseekv32,glm5,qwen3,qwen3coder}, and systems stacks optimize expert-parallel execution~\cite{megablocks,tutel,deepspeedmoe} or orchestrate across CPU-GPU memory tiers~\cite{fiddler,pregatedmoe,powerinfer,flexgen}. All take the per-expert HBM footprint as fixed and schedule around it; FORGE instead eliminates that footprint on replaced layers, collapsing the minimum-fit fleet into a single node and removing cross-node all-to-all from decode.

\noindent\textbf{LLM Serving and KV Cache Management.} Serving frameworks schedule over fixed weight and KV-cache cost~\cite{orca,vllm,sarathiserve,alpaserve,sglang} or disaggregate prefill from decode~\cite{distserve,splitwise,mooncake}, while KV-cache interventions shrink the dynamic state via attention sinks, heavy-hitter eviction, and transform coding~\cite{streamingllm,h2o,turboquant,transformcoding}. These target Attention; FORGE targets the static MoE FFN weights that dominate the remaining decode pressure, and composes with disaggregation by deploying the RGU on decode nodes only.

\noindent\textbf{Near-Data Processing and Transformer Accelerators.} Processing-in-Memory~\cite{samsunghbmpim,aim,neupims,attacc} shortens the weight-movement path without reducing storage; attention accelerators~\cite{spatten,flashattention,flashattention2,flashattention3} and state-space models like Mamba~\cite{mamba} target layers whose volume is dwarfed by MoE FFN weights. FORGE removes the weight data from memory rather than moving computation closer to it, integrating with standard Tensor Core GEMM pipelines~\cite{cutlass,h100wp} without requiring a new memory substrate or a separate accelerator chip.

\section{Conclusion}
\label{sec:conclusion}

FORGE replaces HBM weight loading with procedural on-the-fly generation for Sparse MoE FFN layers. The Rotor Generation Unit (fixed-function accelerator within each SM) executes rotor operations in a dedicated pipelined datapath and writes generated weights directly into Shared Memory, bypassing the HBM hierarchy without contending for SIMT resources. This converts the memory-bandwidth-bound MoE FFN decode into a compute-bound operation at ${\sim}0.9\%$ SM area overhead (0.032~$mm^2$/SM at 7\,nm).

Across five Sparse MoE models with $30\%$ of the MoE FFN layers replaced, we estimate $1.3$--$2.8\times$ higher system throughput at matched GPU count and $1.10$--$1.15\times$ lower per-user decode latency at short context, at a cost in downstream accuracy relative to the same model fine-tuned without rotor replacement (Section~\ref{sec:quality_validation}). Underneath, KV cache capacity grows by $1.2$--$2.6\times$ and the per-expert FP8 kernel speeds up by $2.3$--$3.2\times$ independent of $r$. Higher ratios promise larger system gains; recovery methods that preserve accuracy at those ratios are future work. As MoE expert counts grow, stored weight volume scales linearly while rotor parameter seeds scale logarithmically, so the capacity advantage of procedural generation widens with model scale.

\begin{acks}
We thank the anonymous reviewers for their feedback.
\end{acks}

\bibliographystyle{ACM-Reference-Format}
\balance
\bibliography{main}

\end{document}